\documentclass[
               reprint, twocolumn,
               preprintnumbers,
               amsmath,
               amssymb,
               showpacs,
               aip,jcp,
               superscriptaddress]{revtex4-1}

 \usepackage{ucs}
 \usepackage[utf8x]{inputenc}
 \usepackage{latexsym}
 \usepackage{amsfonts}
 \usepackage{amsmath}
 \usepackage[dvips]{graphicx}
 \usepackage{color}
 \usepackage{bm}
 \usepackage{hyperref}
 \usepackage{lineno}
\graphicspath{ {./fig/} }
\hypersetup{
   colorlinks=true,       
   linkcolor=magenta,          
   citecolor=magenta,        
   filecolor=magenta,      
   urlcolor=blue           
}

\begin{document}
\author{Krishna Reddy Nandipati}
\email[e-mail: ]{krishna.nandipati@pci.uni-heidelberg.de}
\affiliation{Theoretische Chemie,
             Physikalisch-Chemisches Institut,
             Universität Heidelberg,
             Im Neuneheimer Feld 229, 69120 Heidelberg, Germany}

\author{Oriol Vendrell}
\email[e-mail: ]{oriol.vendrell@pci.uni-heidelberg.de}
\affiliation{Theoretische Chemie,
             Physikalisch-Chemisches Institut,
             Universität Heidelberg,
             Im Neuneheimer Feld 229, 69120 Heidelberg, Germany}
\affiliation{Centre for Advanced Materials,
        Universit\"at Heidelberg,
        Im Neuenheimer Feld 205,
        69120 Heidelberg, Germany}



\title{On the Generation of Electronic Ring Currents under Vibronic Coupling Effects}

\date{\today}

\begin{abstract}
    \noindent
    We study the generation of electronic ring currents in the presence of
    nonadiabatic coupling using circularly polarized light. For this, we
    introduce a solvable model consisting of an electron and a nucleus rotating
    around a common center and subject to their mutual Coulomb interaction. The
    simplicity of the model brings to the forefront the non-trivial properties
    of electronic ring currents in the presence of coupling to the nuclear
    coordinates and enables the characterization of various limiting situations
    transparently. Employing this model, we show that vibronic coupling effects
    play a crucial role even when a single $E$ degenerate eigenstate of the
    system supports the current. The maximum current of a degenerate eigenstate
    depends on the strength of the nonadiabatic interactions. In the limit of
    large nuclear to electronic masses, in which the Born-Oppenheimer
    approximation becomes exact, constant ring currents and time-averaged
    oscillatory currents necessarily vanish.
\end{abstract}

\maketitle

\section{Introduction}

Electronic currents triggered by circularly polarized light in ring-shaped
molecules~\cite{barth2006unidirectional,
    barth2006periodic,%
    barth2007electric,%
    barth2010quantum,%
    ulusoy2011correlated,%
    mineo2016induction,%
    mineo2017quantum1,%
    mineo2017quantum2,%
    liu2018attosecond,%
    kanno2018laser}
and
materials~\cite{per05:245331,%
    matos2005photoinduced,%
    rae07:157404}
have become a subject of much interest due to their
potential applications in optoelectronics~\cite{anthony2006functionalized},
e.g. as a platform for fast
switching qubits~\cite{rae07:157404}.
Recent advances in femto- and attosecond laser
technology~\cite{bandrauk2004attosecond,krausz2009attosecond} have motivated
theoretical investigations on the generation of electronic ring currents in
time-scales comparable to the fastest vibrational dynamics in
molecules~\cite{yuan2017attosecond,liu2018attosecond,kanno2018laser}.
It is now well established that ultrashort laser pulses with a sufficient
bandwidth can trigger the migration of electronic charge across molecular
structures in attosecond time-scales~\cite{kul05:10,rem06:6793,kra15:790,cal14:336}.
Ultrafast charge migration implies the superposition of electronic states with
relatively large energy gaps between them, which often feature very different
potential energy surfaces.
Hence, after the electrons are set in
motion, nuclei quickly respond to
the electronic excitation and inevitably lead to
decoherence~\cite{vac17:83001,arn17:33425}
and
electronic energy relaxation through vibronic coupling
effects~\cite{tim14:113003,des15:426}.
On the other hand, circular ring currents require the presence of doubly
degenerate $E$ electronic components, present in molecular systems with at least
a 3-fold or higher symmetry axis, and one may be lead to believe that the
coupling between electrons and nuclei and the corresponding vibronic coupling
effects will not substantially affect the laser-triggered electronic
circulation in these systems.

%
Up to now, most theoretical studies on the generation and control of
ring currents by applying circularly polarized laser fields have
not considered vibronic coupling effects~\cite{barth2006unidirectional,barth2006periodic,barth2007electric,%
barth2010quantum,ulusoy2011correlated,mineo2016induction,%
mineo2017quantum1,mineo2017quantum2,liu2018attosecond,per05:245331,%
    matos2005photoinduced,%
    rae07:157404}.
The relevant time-scales for these currents are at least tens of femtoseconds.
Therefore, it is inevitable that vibronic effects will
influence to some extent the electronic dynamics~\cite{tim14:113003,des15:426}: it has
been reported that the
build up of strong correlations between electrons and nuclei can take as little as 1 to
2~fs in the vicinity of a conical intersection~\cite{li13:038302}.
Previous studies by Kanno \emph{et. al.}
considering
vibronic effects on electronic currents focused on systems without actual
state degeneracy and instead described currents triggered by linearly polarized
light in chiral systems~\cite{kanno2010nonadiabatic,kanno2012laser2,kanno2018laser}.
In these systems, the currents achieved are oscillatory because of
the different energy of the eigenstates involved.

%
The purpose of the present work is to unravel the specific role of vibronic
coupling effects in the generation of sustained ring currents by circularly polarized
light in systems with
two degenerate circulation directions. Owing to their stationary nature,
constant currents must be necessarily supported by a
very small number of energy eigenstates of the
complete molecular Hamiltonian,
even only one, which belong to degenerate pairs.
Even in cases where only one eigenstate is involved, we will discuss how
vibronic effects can be very significant and can lead to an almost complete
suppression of the ring current. We will also establish the general requirements
and laser pulse properties needed in order to achieve either stable and
sustained currents, or highly oscillatory ones.
%
To this end, we consider a model system consisting of two concentric rings, the
innermost one constraining the motion of a positively charged atomic nucleus, and
the outermost one constraining the motion of an electron around a common center.
The two rings lie on the $(x,y)$-plane and the laser field is assumed to
propagate perpendicular to the plane of the particles along the $z$-direction.
Despite its simplicity, this ring-ring-model (RRM) fulfills two key
characteristics: first, it captures the fundamental features of molecular
systems that are able to present circular electronic currents with two
degenerate rotation directions and second, it captures the effect of the
coupling to the nuclei.
On the one hand, the RRM consists of a rotational axis of at least third order,
resulting in doubly degenerate \emph{vibronic} states of an $E$ symmetry
representation within the corresponding point group. On the other hand, it
features a nuclear coordinate that breaks this symmetry for the electronic
subsystem if fixed at an arbitrary point in nuclear configuration space.
The model represents a class of vibronic coupling phenomena resulting from
either Renner-Teller (RT)~\cite{koppel1981theorylinearmolecules} or Jahn-Teller
(JT)~\cite{Longuet-Higgins} effects, in which the nuclear motion of appropriate
symmetry couples the doubly degenerate electronic states of $E$ symmetry. 
The RRM including an external laser field is considered numerically exactly
without invoking the BO approximation. Moreover, several of its important
features can be understood based on analytical considerations.
We also show how the Hamiltonian of the RRM can be expressed as a vibronic
coupling Hamiltonian (VCH)~\cite{koppel} in a diabatic electronic basis, which
is an exact representation as well.
This fact underpins an important observation, namely that a VCH provides the
necessary theoretical framework to describe electronic ring currents under
nonadiabatic couplings between electrons and nuclei.

%


\section{Theory}

%
In the following we introduce the RRM system and some theoretical
considerations.
The electronic and nuclear motion are constrained to two concentric rings
of fixed radii $r$ and $R$, respectively, lying on the $(x,y)$-plane.
%
The electronic and nuclear angular degrees of freedom are denoted by
$\theta$ and $\alpha$, respectively.
%
The total Hamiltonian including the light-matter interaction
term within the semiclassical dipole-approximation \cite{shapiro2003principles}
reads
\begin{align}\label{H}
    \hat{H} & = \hat{H}_\textrm{mol} -\bm{\mu} \cdot \mathbf{E}(t) \\\nonumber
            & = \hat{T}_{\alpha} +
                \hat{T}_{\theta} +
                \hat{V}(\alpha, \theta)
                -\bm{\mu} \cdot \mathbf{E}(t)
\end{align}
with kinetic energy (KE) terms
\begin{equation}
\label{T}
  \hat{T}_{\gamma} = -\frac{\hbar^{2}}{2I_{\gamma}}
    \frac{\partial^2}{\partial\gamma^2} \;\;\;\; \gamma\to(\alpha,\theta)
\end{equation}
  and moment of inertia $I_\gamma = m_\gamma r_\gamma^2$ with $r_\gamma \to (R,r)$ and  $m_\gamma \to (M,m)$. Here $M$ and $m$ are the mass of the nucleus and electron, respectively.
  The Coulomb interaction of the electron and the nucleus is then given by
  \begin{align}\label{V1}
      \hat{V}(\alpha, \theta) & =
      \frac{-Q e^2}{\sqrt{
          (r\cos{\theta}-R\cos{\alpha})^2 +
          (r\sin{\theta}-R\sin{\alpha})^2}
      } \\\nonumber
      & =
      \frac{-Q e^2}
           {\sqrt{
              (r^2 + R^2 -2 r R \cos(\theta - \alpha)
           }
          } \\\nonumber
      & =
      \frac{-Q e^2}
           {\sqrt{
              (A - B \cos(\theta - \alpha)
           }
          },
  \end{align}
  where $Q$ is the charge of the nucleus and the $A$ and $B$ constants are
  implicitly defined.
  Although the 2D Coulomb interaction can be handled numerically
  in the solution of the time-dependent Schrödinger equation (TDSE), we
  introduce here a second order approximation to the interaction similar in
  spirit to a multipolar expansion, namely the potential is expanded to second
  order around $\cos(\theta - \alpha)=0$. This approximation is very good as long
  as the radius of the outer ring is several times larger than the inner ring,
  as in our case, and leads to the interaction potential
\begin{equation}\label{V2}
    \hat{V}(\alpha, \theta) =
    v_0
    - \frac{1}{2}\kappa_1\cos(\theta - \alpha)
    - \frac{1}{4}\kappa_2\cos^2(\theta - \alpha).
\end{equation}
  with
  $v_0=-Qe^2/\sqrt{A}$,
  $\kappa_1=Qe^2 B/A^{3/2}$ and
  $\kappa_2=3 Qe^2 B^2/2A^{5/2}$.
  The first term $v_0$ is the constant charge-charge
  interaction, which is subtracted from the potential in the following.
  The term proportional to $\cos(\theta-\alpha)$
  couples the doubly degenerate electronic states with the
  nondegenerate ground electronic state,
  whereas the term proportional to $\cos^2(\theta-\alpha)$
  couples directly the
  doubly degenerate electronic states.
  This expansion of the Coulomb interaction facilitates the analytical
  description of the electron-nucleus interaction terms and of the angular
  momentum transfer among the two subsystems.
  In the numerical simulations of the RRM system the parameters are
  (
  $r=20$,
  $R=1$,
  $m=1$,
  $M=2000$,
  $Q=1$),
  all in atomic units, unless otherwise specified.

  We introduce now a Born-Huang (BH) expansion \cite{born1954dynamical}
  of the
  nuclear-electronic wave function $\psi(\alpha, \theta)$ as is customary in the
  description of nonadiabatic effects in molecular systems.
  A convenient diabatic electronic basis are the
  eigenstates
  of the uncoupled electron, $\hat{T}_\theta$, 
  given by $\varphi_l(\theta)=e^{il\theta}/\sqrt{2\pi}$,
  resulting in the wave function
  \begin{equation}\label{BH}
    \psi(\alpha, \theta, t) = \sum_{l=-L}^{L}
  \chi_l(\alpha, t) \varphi_l(\theta) = \sum_{l=-L}^{L} \chi_l(\alpha,t)
  \left\{\frac{1}{\sqrt{2 \pi}}e^{i l \theta}\right\},
  \end{equation}
  where the population of each electronic state follows from the norm-squared of
  the nuclear amplitude
  $P_l(t) = \langle \chi_l(t) | \chi_l(t)\rangle$.
  In the discussion below it will be sometimes useful to expand the nuclear
  wavepackets in the basis of eigenstates of the uncoupled nuclear coordinate,
  \mbox{$\chi_l(\alpha,t)$ = $\frac{1}{\sqrt{2 \pi}}\sum_{k=-\nu}^{\nu} a^{l}_k(t) e^{i k
  \alpha}$}.
  The matrix representation of the vibronic coupling Hamiltonian (VCH)
  with potential matrix elements
  $\langle\varphi_{l^\prime}|(\hat{H}_\textrm{mol} - \hat{T}_\alpha)
  |\varphi_l\rangle$
  for $L=1$, i.e. in the space of electronic eigenstates with quantum numbers
  $l=\{-1,0,1\}$, reads
\begin{equation}\label{HV}
\bm{H} =
\hat{T}_{\alpha}{\bm{I}_{3 \times 3}}
+
  \left(
    {\begin{array}{ccc}
            \frac{\hbar^2}{2I_\theta}   &   \frac{\kappa_1}{4} e^{i\alpha} & \frac{\kappa_2}{16} e^{2i\alpha} \\
          \frac{\kappa_1}{4} e^{-i\alpha} &  0           &  \frac{\kappa_1}{4} e^{i\alpha}           \\
         \frac{\kappa_2}{16} e^{-2i\alpha}&   \frac{\kappa_1}{4} e^{-i\alpha}         & \frac{\hbar^2}{2I_\theta}
    \end{array}}
\right)
\end{equation}
where \textbf{\textit{I}}$_{3 \times 3}$  is the 3 $\times$ 3 unit matrix.
The resulting coupling scheme in the VCH follows transparently. The diabatic
electronic state with $l=0$ couples with the $l=1$ and $l=-1$ states in first
order, and this interaction changes the angular momentum of the nuclear
coordinate $\alpha$ by $\pm\hbar$.  On the other hand, the $l=+1$ and $l=-1$
states are directly coupled among each other in second order which changes the
angular momentum of the nuclear coordinate $\alpha$ by $\pm2\hbar$.
The present model is similar in spirit to that considered by
Longuet-Higgins~\cite{Longuet-Higgins} except for the fact that our model
has an interaction potential with a periodicity of $2\pi$ from the perspective
of the electronic coordinate,
whereas that of Longuet-Higgins has a periodicity of
$\pi$, namely it is proportional to $\cos(2\theta - \alpha)$.
The $\pi$ periodicity of the Longuet-Higgins model results in the direct
interaction between the degenerate electronic $E$ components already in first
order, which corresponds to JT-type coupling~\cite{Longuet-Higgins}.
In the taxonomy of JT and RT
models, the RRM resembles a RT Hamiltonian, characterized by $2^{nd}$-order coupling terms, featuring both
direct and indirect coupling of the degenerate electronic components. This fact
makes it amenable to comparisons with general vibronic coupling situations in
molecules.

The VCH in Eq.~(\ref{HV}) provides a complete representation of the
electronic-nuclear coupling of the system
within the selected group of electronic states and
under the coupling potential in Eq.~(\ref{V2}).
It has a similar structure compared to a molecular VCH with a $C_n$
symmetry axis of at least order $n=3$, where always doubly degenerate states of
an $E$ symmetry representation are present \cite{bersuker2006jahn}.
Only the nuclear dependency of the coupling terms has a simpler structure than in
molecular cases, where for example other coupling and tuning nuclear coordinates
would be present~\cite{koppel}.
In the same way as the real-valued $p_x$ and $p_y$ orbitals of an atom, the real
$\varphi^{E}_x$ and $\varphi^{E}_y$ electronic states of a molecular system can
be combined to yield electronic states $\varphi^{E}_\pm \propto \varphi^{E}_x
\pm i \varphi^{E}_y$, featuring a net current; these are analogous to the $l=+1$
and $l=-1$ diabatic electronic states of the RRM model. 

  It is convenient in the following to introduce as a basis the product of
  electronic and nuclear angular momentum eigenstates $|k,l\rangle$ (with
  $\langle \alpha, \theta|k,l\rangle$= $\frac{1}{2\pi}e^{ik\alpha}
  e^{il\theta}$), where $k$ and $l$ are the angular momentum quantum numbers
  of the electronic and nuclear degrees of freedom, respectively.
  The coupling between pairs of $|k,l\rangle$ basis states follows from simple
  integration of these with the interaction potential in Eq.~(\ref{V2}):
  \begin{widetext}
  \begin{eqnarray}\label{NACME2}
    \langle k^{'},l^{'}|\hat{V}^{(m)}(\alpha,\theta)|k,l\rangle &=& \beta_m
    \left(1-S(k-k^{'}) S(l-l^{'})\right)
    \delta_{|k-k^{'}|,m}\delta_{|l-l^{'}|,m},
  \end{eqnarray}
  \end{widetext}
  where $m$=1, 2 and $\beta_1$= $\kappa_1/4$; $\beta_2$= $\kappa_2/16$ and the
  function $S(z)$ returns the sign of its real argument modulo one.
  The presence of the sign functions $S(k-k^{'})$ and $S(l-l^{'})$ restricts the
  coupling to the cases in which the angular momentum
  quantum number of electron is increased by $\hbar m$ while that of nucleus correspondingly is decreased by $\hbar m$, or
  vice versa.
  Therefore it is meaningful to introduce the quantum number $q=k+l$, which is
  conserved by the vibronic coupling, and to
  note that the total Hamiltonian separates into blocks according to the value
  of $q$.
The last term in Eq.~(\ref{H}) describes the matter-light interaction with dipole \mbox{$\bm{\mu} = -r(\bm{\epsilon}_x\cos(\theta)+\bm{\epsilon}_y\sin(\theta))$}
(the nuclear contribution to the dipole can be safely neglected because
the fields applied are not resonant with direct transitions in the nuclear
coordinate) and electric field
\mbox{$\mathbf{E}(t) =\bm{\epsilon}_x E_x(t) +  \bm{\epsilon}_y E_y(t)$}
terms, where $\bm{\epsilon}_u$
is the unitary polarization vector pointing in the $u$-direction.
The electric field of the laser pulses $E_u(t)$ is derived from the vector
potential
\begin{equation}\label{vectorPot}
    A_u(t) = \frac{E_0}{\omega}S(t)\sin({\omega t -\phi_u})
\end{equation}
as $E_u(t)=-\partial A_u(t) / \partial t$, where we use a sine-squared envelope
function
\mbox{$S(t) = \tilde{\Theta}(t-\tau)\sin^2 \left( \frac{\pi t}{\tau} \right)$}.
$E_0$, $\tau$ and $\omega$ are the maximum amplitude, pulse duration (start to
end) and carrier
frequency of the pulse, respectively, and $\tilde{\Theta}(t-\tau)$ is the
inverse Heaviside step function.
We take all pulse parameters for both $(x,y)$ polarization directions to be equal
except for the $\phi_x$ and $\phi_y$ phases.
A right circularly polarized pulse (RCP) corresponds to the phases
$(\phi_x=\pi/2, \phi_y=0)$
and
a left circularly polarized pulse (LCP) to the phases
$(\phi_x=0, \phi_y=\pi/2)$.
\subsection{Ring currents under vibronic coupling effects}
The electronic component $J^{(\theta)}(\alpha, \theta,t)$ of the total wave function current
at coordinates $(\alpha, \theta)$
follows from the continuity condition of the probability density~\cite{Messiah1999}
and is given by
\begin{equation}\label{J1}
 J^{(\theta)}(\alpha,\theta,t) = \frac{1}{{I_\theta}}Re
    \{\psi^{*}(\alpha,\theta,t) \hat{l}_\theta \psi(\alpha,\theta,t)\}, 
\end{equation}
where $\hat{l}_\theta$= $-i\hbar {\partial_\theta}$.
The electronic component of the current as a function of the nuclear
coordinate(s) is of not much practical use. In actual applications, one needs instead
the electronic current averaged over the nuclear coordinates, which in
the present model reads
\begin{equation}\label{J11}
 J_{e}(\theta,t) = \frac{1}{I_\theta 2\pi} \int_{0}^{2\pi}
    J^{(\theta)}(\alpha,\theta,t) d\alpha.
\end{equation}
 \noindent
Using the Born-Huang expansion, Eq.~(\ref{BH}), with Eqs.~(\ref{J1})
and (\ref{J11}), one arrives at
the following expression for the nuclear-averaged electronic current:
\begin{widetext}
 \begin{equation}\label{J2}
 J_{e}(\theta,t) = \frac{1}{I_\theta 2\pi} \int_{0}^{2\pi}
     \left[\sum_{l=-L}^{L} \hbar l |\chi_l(\alpha,t)|^2
     + \Re \left\{\sum_{l^\prime = -L}^{L}
     \sum_{l=-L}^{L}(1-\delta_{l^\prime l}) \hbar l \chi^{*}_{l^\prime}(\alpha,t)
     \chi_l(\alpha,t) e^{i (l-l^\prime) \theta} \right\}\right] d\alpha.
\end{equation}
\end{widetext}
The first term in the above equation is $\theta$-independent and describes the
contribution of the $l$-th electronic state to the overall ring current.  The
second term is a fluctuating contribution originating from the interference
between the $l$-th and $l^{\prime}$-th electronic basis states at a given nuclear geometry $\alpha$.
This fluctuating component along $\theta$ vanishes when integrating
$J_{e}(\theta,t)$ over the closed path of the \emph{electronic coordinate}
due to the imaginary
exponential term.
Thus, averaging over the nuclear coordinate and integrating over the
electronic closed path one
arrives at the expression for the net time-dependent electronic ring current
\begin{equation}\label{J21}
    J_{e}(t) = \frac{1}{4\pi^2 I_\theta} \int_{0}^{2\pi}
    J_{e}(\theta,t) d\theta = \frac{\hbar}{4\pi^2 I_\theta}
    \sum_{l=-L}^{L}
    l P_l(t).
\end{equation}
 where $P_l(t)$ = $\int_{0}^{2\pi} |\chi_l(\alpha,t)|^2 d\alpha$.
 This expression for the net ring current tells that the contribution of the
 $l$-th electronic-current state is proportional to the corresponding
 population, namely the nuclear coefficient squared in the BH wave function
 expansion.
 This simple expression is reached because our electronic diabatic basis consists,
 for convenience, of complex-valued current eigenstates instead of their
 real-valued linear combinations.
 In a molecular calculation one would introduce the corresponding
 $\varphi_\pm^E$ diabatic electronic-current states
 to arrive at an analogous expression for the net ring currents.

 Equation~(\ref{J21}) illustrates how the creation of a stable imbalance of the
 population (the norm of the nuclear wavepackets) of the current components
 corresponding to
 opposite directions (i.e. $l=\pm 1$ in the model) results in an electronic ring
 current.
 If the population imbalance is stationary, so is the corresponding current.
 If, on the other hand, the electronic-current state populations vary, which can \emph{only} be
 the result of either nonadiabatic couplings or interactions with
 an external electromagnetic field, the net electronic ring current
 $J_e(t)$ varies in time as well.



  The coupling terms in Eq.~(\ref{NACME2}) illustrate
  the basic mechanism of the
  electronic-nuclear interactions affecting ring currents, which consists of
  the exchange of angular momentum between the electron and the nucleus.
  According to Eq.~(\ref{NACME2}) and restricting the electronic subspace to $l=\{-1,0,1\}$ as above, the basis state $|k, 0\rangle$ couples only to the states $|k+1, -1\rangle$ and $|k-1, +1\rangle$.
  The three basis states $|0,0\rangle$, $|1,-1\rangle$ and
  $|-1,1\rangle$ result in a $3\times 3$ Hamiltonian matrix with $q=0$ that upon
  diagonalization yields three non-degenerate vibronic eigenstates (eigenstates
  of the complete nuclear-electronic Hamiltonian)
  \begin{equation}
      \label{BHCVWFnd}
      |{j}_0\rangle =
              c^{j_0}_{-1, 1} |-1,  1\rangle
            + c^{j_0}_{ 0, 0} | 0,  0\rangle
            + c^{j_0}_{ 1,-1} | 1, -1\rangle,
 \end{equation}
 where the notation  $|{j}_{q}\rangle$ is used for the vibronic states.
 The expansion coefficients obey $|c^{j_0}_{-1, 1}| = |c^{j_0}_{1, -1}|$ on symmetry
 grounds.
 Therefore, not surprisingly vibronic eigenstates with $q=0$
 feature no net electronic current.
 The operation of reversing the sense of circulation of all particles,
 ($k\to-k$, $l\to-l$), is closed within the basis states, and the corresponding
 eigenstates belong to a completely symmetric representation.


  On the other hand, the doubly degenerate vibronic states originate from
  the two blocks with quantum numbers $\pm q$ and $|q| \geq 1$.
  %
  One such block,
  e.g.,
  $|k,0\rangle$, $|k+1, -1\rangle$ and $|k-1, +1\rangle$ with $q=1$, results
  in the corresponding $3\times 3$ Hamiltonian matrix which can be diagonalized
  to yield the three non-degenerate states
  \begin{equation}\label{BHCVWFdd}
    |{j_1}\rangle =
        c^{j_1}_{k-1, 1}  |k-1, 1\rangle
      + c^{j_1}_{k, 0}    |k, 0\rangle
      + c^{j_1}_{k+1, -1} |k+1, -1\rangle.
  \end{equation}
  These vibronic eigenstates feature an electronic ring current
  because, in general, $|c^{j_q}_{k-1, 1}| \neq |c^{j_q}_{k+1, -1}|$ and
  therefore they support an imbalance of the contribution to each
  electronic current circulation
  direction.

  Inverting the sense of circulation of all particles, $q\to -q$,
  results in the basis
  functions
  $|-k,0\rangle$, $|-k+1, -1\rangle$ and $|-k-1, +1\rangle$ with $q=-1$.
  The diagonalization of the
  corresponding Hamiltonian matrix yields the triad of
  $|{j_{-1}}\rangle$ states, degenerate one to one with the corresponding
  $|{j_1}\rangle$ eigenstates.
  Finally, the expansion coefficients of the vibronic eigenstates
  $|{j_1}\rangle$ and $|{j_{-1}}\rangle$
  are related by
  $c^{j_1}_{k \pm 1, \mp 1}=c^{j_{-1}}_{-k \mp 1, \pm 1}$
  and
  $c^{j_{1}}_{k, 0}=c^{j_{-1}}_{-k, 0}$.


  We can already see through the general strategy to generate a ring current
  under nonadiabatic electronic-nuclear couplings.
  Stated in terms of eigenstates of the complete Hamiltonian, it will consist in
  creating a population imbalance within pairs of vibronic (not electronic)
  states $|{j_{q}}\rangle$ and $|{j_{-q}}\rangle$. If each $(j_q,j_{-q})$
  pair of eigenstates of a given $\pm q$ is equally populated, no net current will
  be achieved. This is because the individual net currents of $j_q$ and $ j_{-q}$ are equal and opposite, as $c^{j_q}_{k \pm 1, \mp 1}=c^{j_{-q}}_{-k \mp 1, \pm 1}$.

  \section{Results and discussion}
  Let us assume that a narrow-band laser pulse
  resonant with the transition between the absolute ground state
  $|0_{0}\rangle$ and
  $|{j_{\pm q}}\rangle$, i.e. with central frequency
  \mbox{$\omega = (E_{j_q} - E_0)/\hbar$}, interacts with the system in its ground state
  and transfers population to this pair of eigenstates only.
  In this case, the vibronic wavepacket after the pulse is
  \begin{equation}\label{VWPSC1}
     |\psi(t)\rangle =
           A_{0}(t) |{0}_0\rangle
         + A_{j_{q}}(t)|j_{q}\rangle
         + A_{j_{-q}}(t)|j_{-q}\rangle,
  \end{equation}
  where $|A_{0}(t)|^2$ and $|A_{j_{\pm q}}(t)|^2$ are the populations of
  the ground and excited vibronic states.
  For a linearly polarized pulse $|A_{j_q}(t)|^2 = |A_{j_{-q}}(t)|^2$ follows,
  whereas a circularly polarized pulse results in a complete imbalance with one
  of the two components receiving zero population. This fact can easily be
  checked analytically and it will be illustrated in the
  numerical results below.
  We can further note that the dipole operator in Eq.~(\ref{H}) changes the
  quantum number of the electron by $\pm 1$, and that the ground state belongs to
  the $q=0$ block. Hence, for weak, perturbative
  pulses interacting with the ground state the currents are restricted to
  the subspaces with $q=\pm 1$. The discussion focuses on this one-photon limit
  case in the following.

  The question we ask is, what are the fundamental characteristics
  of the ring currents that are generated by such narrow-band pulses?
  Using that $P_l(t)=|\langle l|\psi(t)\rangle|^2$ in Eq.~(\ref{J21})
  and assuming that the system is described by the state in Eq.~(\ref{VWPSC1})
  one arrives at the ring current associated with a pair of vibronic states
  \begin{equation}\label{J3}
      J_{j_1}(t) = \frac{\hbar}{4\pi^2 I_\theta}
      \Big(|A_{j_1}(t)|^2 - |A_{j_{-1}}(t)|^2 \Big)Y_{j_1},
 \end{equation}
 where $Y_{j_{1}} =|c^{j_{1}}_{0, 1}|^2 -|c^{j_{1}}_{2, -1}|^2$
 and we have used the
 relation $Y_{j_1}=-Y_{j_{-1}}$ to separate the
 common factor.
 The state-dependent molecular factor $Y_{j_1}$ can take values in the range
 $(-1,1)$ and it corresponds to the imbalance of positive versus negative
 electronic current of the $|{j_1}\rangle$ vibronic eigenstate.
 Its value is uniquely determined by the strength of the vibronic coupling of the system.
 In the complete absence of coupling, right- and left-rotating electronic
 contributions do not contribute simultaneously to the same degenerate
 eigenstate because they are not connected via the vibronic coupling terms,
 Eq.~(\ref{NACME2}), in the Hamiltonian. In this case, $Y_{j_1}=\pm 1$. For
 strong nonadiabatic coupling $Y_{j_1}\to 0$ and only a small current within the
 $j_1$ subspace is possible.
 Summarizing, the generation of a ring current within the degenerate space of
 two vibronic states of the complete Hamiltonian requires two ingredients: that the
 molecular factor $Y_{j_q}$ is significant for the $j_{\pm q}$ states, and that
 an imbalance of the populations $|A_{j_q}(t)|^2$ and $|A_{j_{-q}}(t)|^2$
 is achieved by the circularly polarized nature of the light interacting with
 the system.
 Since, by assumption, only two eigenstates with energy $E_{j_1}$ are
 involved, such a current will be stationary (cf. Eq.~(\ref{J3})) once the laser
 pulse is over.
  Let us consider now the case that the incoming pulse is short and has
  sufficient bandwidth to overlap with transitions to a group of optically
  bright vibronic eigenstates.
  In this case, starting again from Eq.~(\ref{J21}) and assuming that after the laser pulse
  the wave function consists of a linear superposition of all states in the $q\pm 1$ blocks,
  one arrives at a generalization of Eq.~(\ref{J3}),
  \begin{widetext}
  \begin{align}\label{Josc}
      J_e(t)
      & = \frac{\hbar}{4\pi^2 I_\theta}
      \Big[ \sum_{j_q}
             \Big( |A_{j_q}(t)|^2 - |A_{j_{-q}}|^2 \Big) Y_{j_q} +
      \Big( 2\Re\{ A_{j_q}^*(t) A_{j_q^\prime}(t) \}
           -2\Re\{ A_{j_{-q}}^*(t) A_{j_{-q}^\prime}(t) \} \Big) Y_{j_q j_q^\prime} \Big].
  \end{align}
  \end{widetext}
  A derivation of the most general expression without the limitation to the
  $q=\pm 1$ blocks is given in the supporting information (SI).
  After the pulse is over, oscillatory contributions to the ring current arise
  from interferences within each $q=\pm 1$ branch in the second term of the
  expression.
  Clearly, no quantum interference between the different $q$ blocks resulting in
  current oscillations takes place. Moreover, a linearly polarized pulse creates
  the same populations \emph{and phases} in the two $q$ blocks and the terms in
  parenthesis vanish in the same as discussed for the narrow-band pulse.
  A short, circularly polarized pulse, instead, creates a superposition of states
  of either the $q=1$ or $q=-1$ branches, resulting in an oscillatory ring
  current.
  The molecular factor in the second term is now $Y_{j_q j_q^\prime} =
  ({c_{0,1}^{j_q}})^* c_{0,1}^{j_q^\prime} - ({c_{2,-1}^{j_q}})^*
  c_{2,-1}^{j_q^\prime}$, the generalization to two states of the product of the
  coefficients for each current direction.

  In order to generate net electronic currents according to
  Eq.~(\ref{Josc}), the task is to identify and target
  specific degenerate vibronic eigenstates that are characterized by a net
  imbalance of the electronic current in each direction.
  For molecules with a $C_3$ axis, this means to identify bright degenerate
  vibronic states $|j_{\pm}\rangle$ for which the molecular factor
  $Y_j = |\langle E_+ | j_+\rangle|^2 - |\langle E_- | j_+\rangle |^2$ is as close to 1
  (or -1)
  as possible. Here the $E_\pm$ diabatic electronic states play exactly the same
  role as the $l=\pm 1$ electronic states in the RRM.

\begin{figure}
  \includegraphics[width=8.5cm]{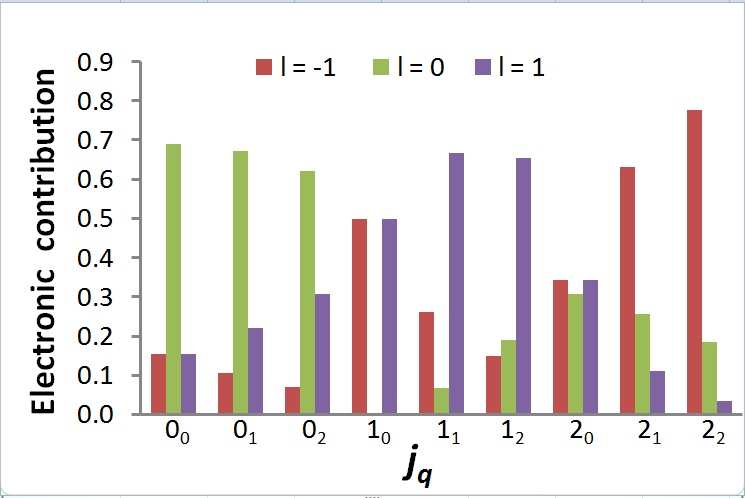}
  \caption {\label{elcontr} Contributions of the  electronic ground ($l=0$) and
     the electronic excited ($l=-1$ and $l=1$) states for the first few vibronic
     states. The transition dipole moment $|\mu_{0_0 j_q}|^2$ is non-vanishing for $j_q$=0$_1$, 1$_1$
  and 2$_1$ }
  

  \end{figure}

  Figure~\ref{elcontr} provides the $l=\{-1,0,1\}$ contributions for the ground
  state and first eight excited states $|j_q\rangle$. For degenerate states, only
  the $q=(1,2)$ branches are shown.
  The molecular factor for each state follows from subtracting the right
  (purple) from the left (magenta) columns.
  Only the states $0_{\pm 1}$, $1_{\pm 1}$ and $2_{\pm 1}$ have
  a non-zero transition-dipole moment ($|\mu_{0_0 j_q}|^2$) with the $0_0$ ground state.
  States $1_1$ and $2_1$ seem both good candidates to generate a ring current by
  applying a RCP pulse. Recall that RCP and LCP pulses couple the ground state
  to the $q=+1$ and $q=-1$ subspaces, respectively.
  The molecular factor for $1_1$ is about $0.4$ for the considered
  model parameters, whereas it is about $-0.52$ for the $2_1$ state. This results in
  the counterintuitive fact that a RCP resonant with the $0_0\to 2_1$ transition
  creates a left-circulating electronic current.
  This is a consequence of the electronic-nuclear coupling. The basis state
  $|2,-1\rangle$ has a larger weight than the basis state $|0,1\rangle$ in
  the vibronic state $|2_1\rangle$. However, this transition has a small Franck-Condon factor
  because the main basis contributions to each eigenstate cancel:
  $\langle 0_0 | \mu_e | 2_1\rangle \approx \langle 0,0|\mu_e|2,-1\rangle
  =\langle 0|2\rangle \langle 0 |\mu_e|-1\rangle = 0$.
  Therefore, in what follows, we study the situation in which a RCP
  laser pulse is resonant with the $0_0\to 1_1$ transition.

  \begin{figure}
  \centering
  \includegraphics[width=8.5cm]{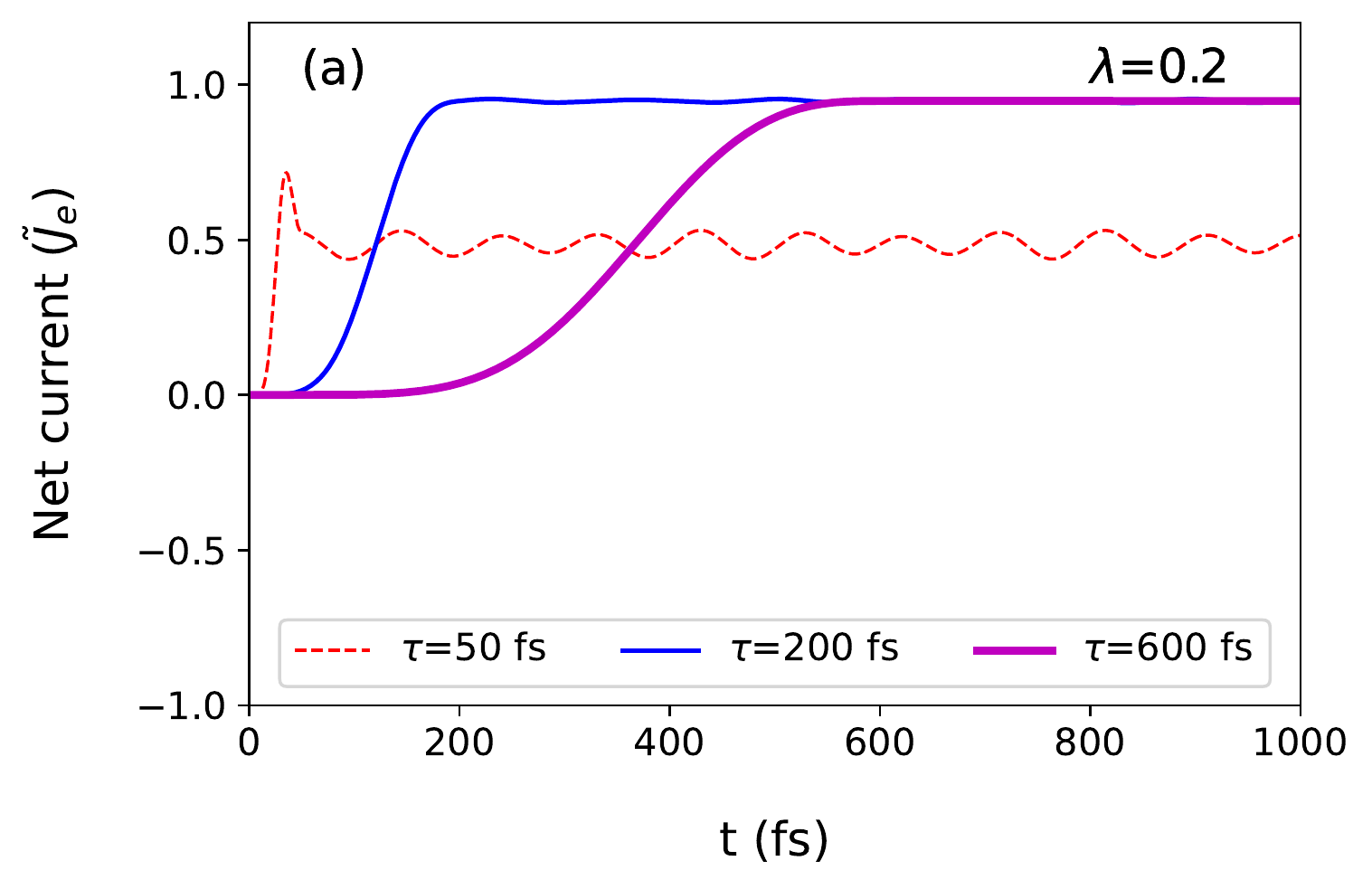}
  \includegraphics[width=8.5cm]{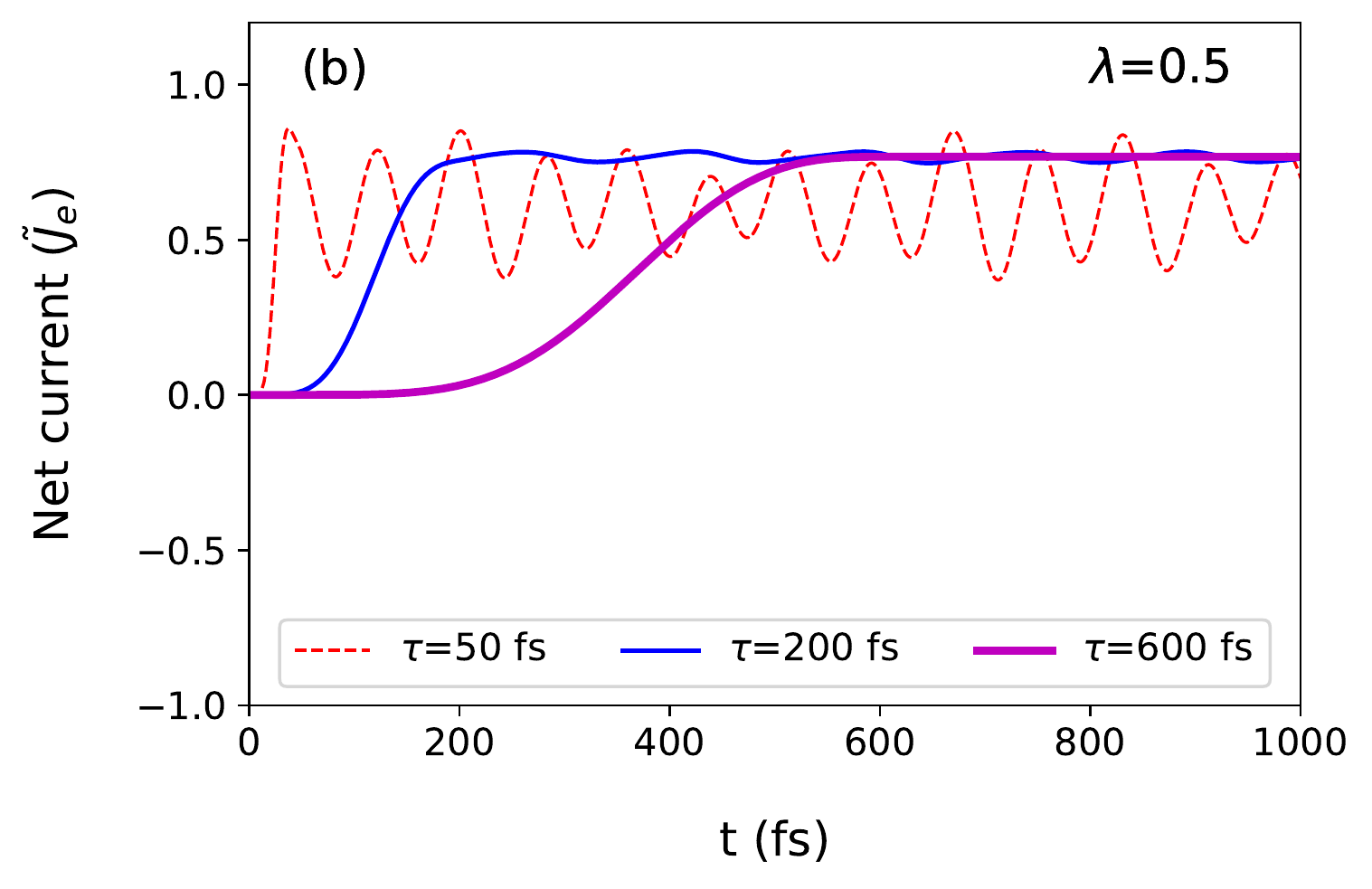}
  \includegraphics[width=8.5cm]{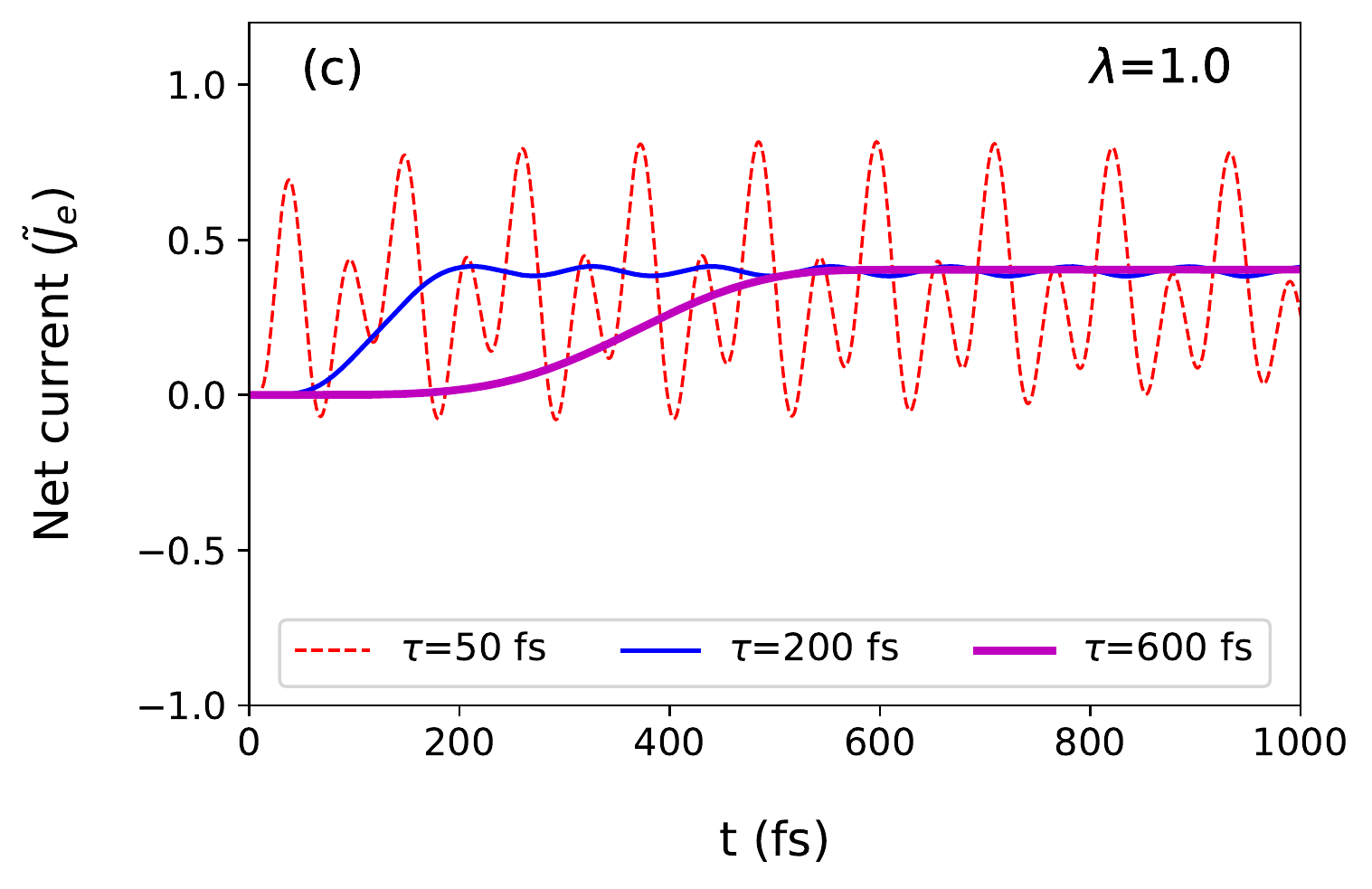}
  \caption{\label{current_l} The normalized net currents ($\tilde{J}_e$) at
  coupling strengths $\lambda$=0.2 (a), $\lambda$=0.5 (b) and $\lambda$=1.0 (c),
  generated by the action of RCP with
  different durations ($\tau$), as function of time. 
  }
  \end{figure}

  Figure~\ref{current_l} shows the current generated by pulses of various
  durations, ranging from 50 to 600~fs. Instead of the net current, we plot the
  quantity $\tilde{J}_e(t) = J_e(t)/\gamma (1-P_0(\tau))$,
  where $\gamma=\hbar/4\pi^2 I_\theta$
  and $P_0(\tau)$ is the population of the ground state after the pulse is over.
  This normalized net current can take values between $(-1,1)$ and its
  introduction serves two purposes.
  First, the normalization by the amount of excitation induced by the laser
  makes the results independent of the laser intensity in the perturbative
  regime. We have lowered the laser intensity to an amount in which the
  \emph{normalized} current becomes independent of the laser intensity.
  Second, it pinpoints the fact that not all the population transferred to a
  vibronic eigenstate contributes to the net current. This is limited by the
  molecular factor. For example, a molecular factor of $0.4$ sets the limit of
  the normalized current associated with this eigenstate to $0.4$.
 \begin{figure*} \centering
      \includegraphics[width=8.5cm]{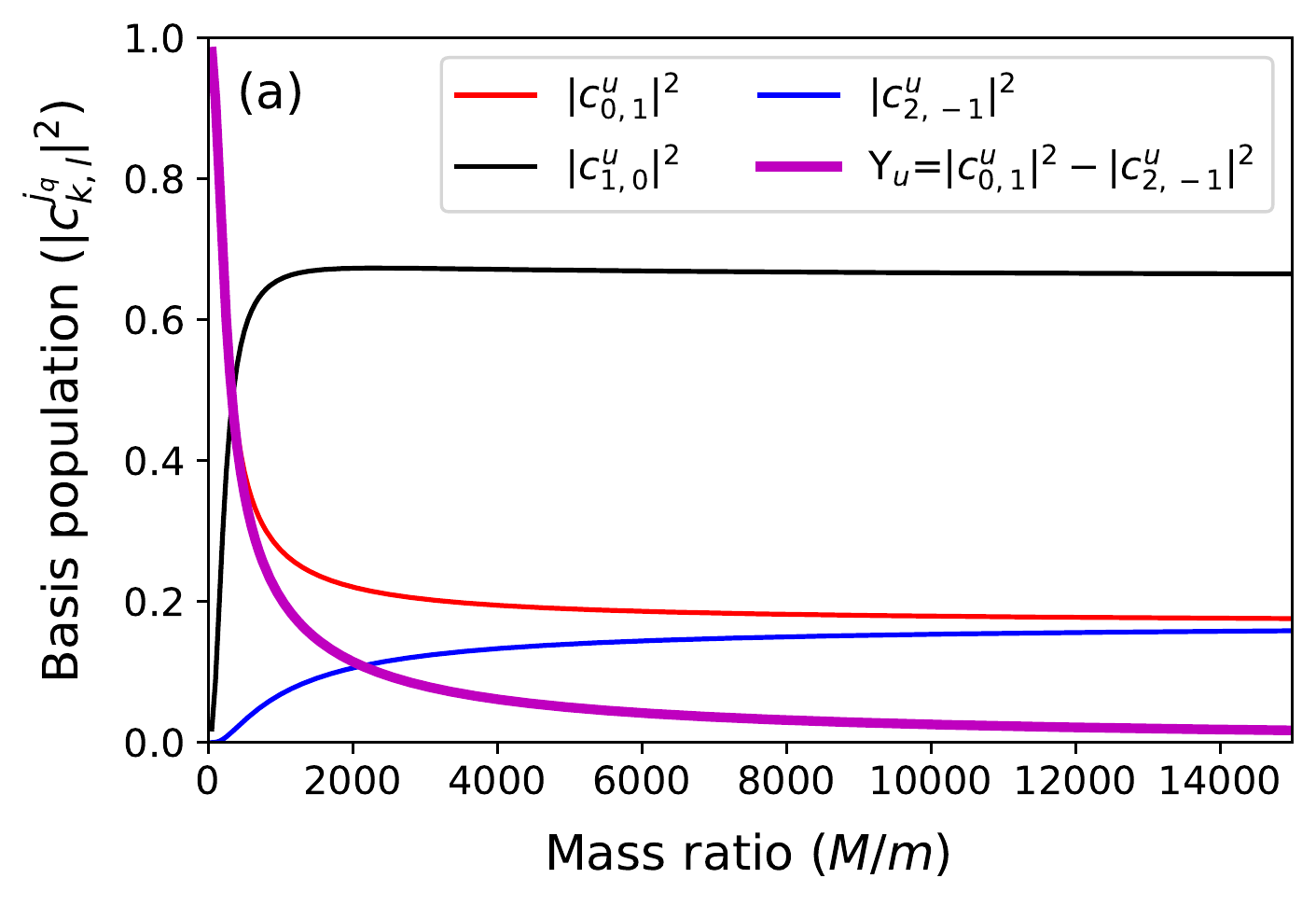}
      \includegraphics[width=8.5cm]{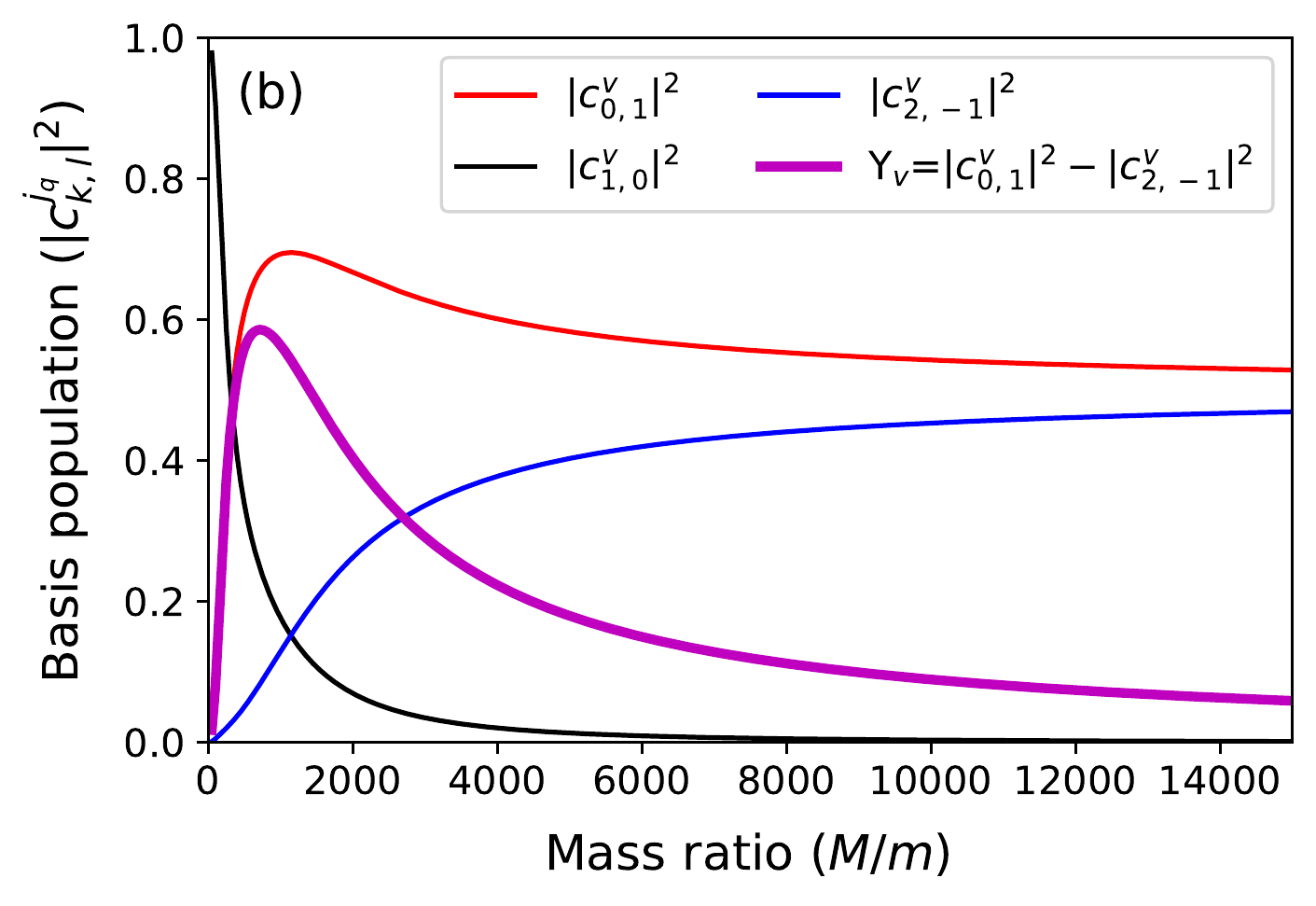} \\
      \includegraphics[width=8.5cm]{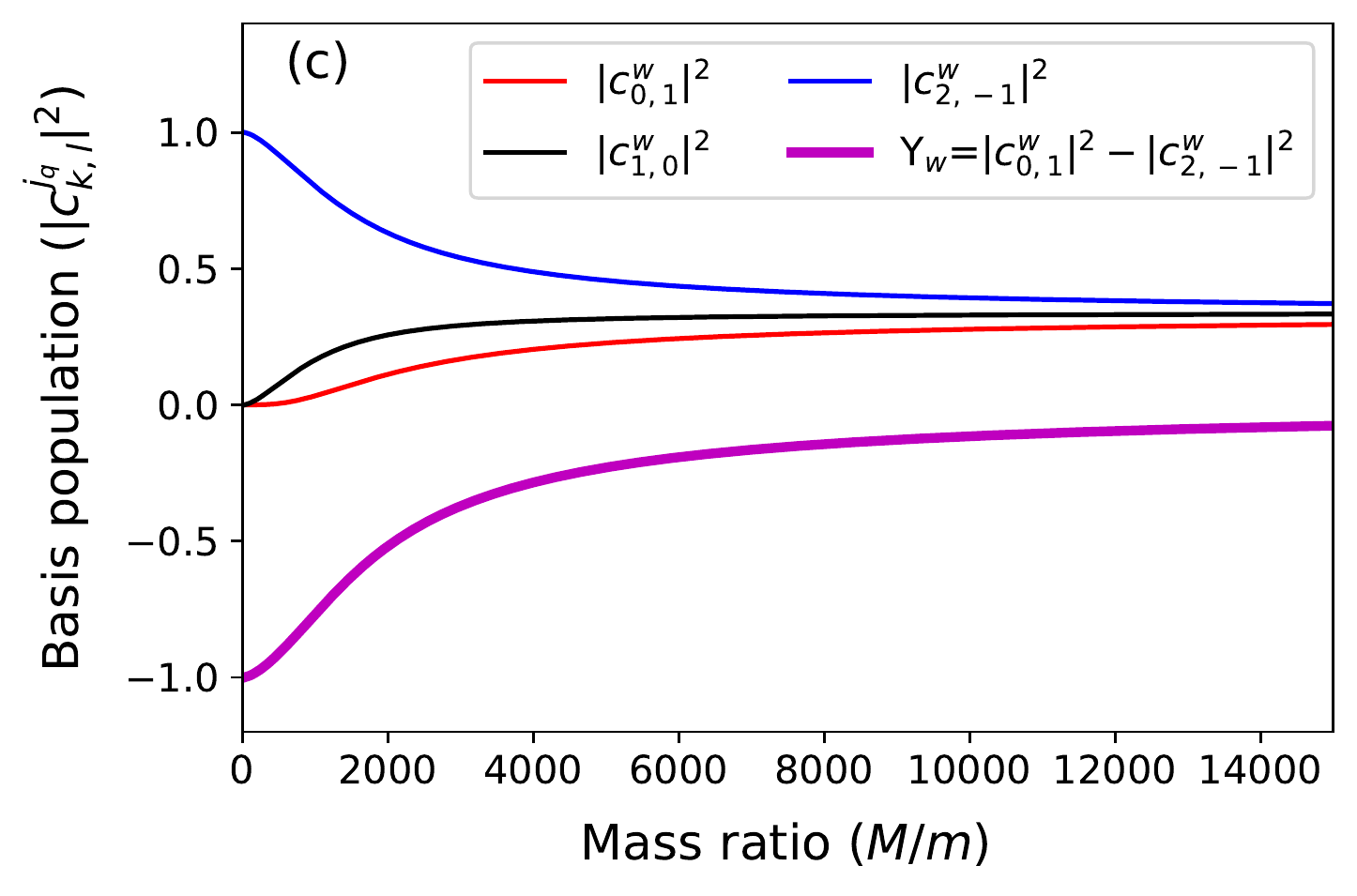}
      \includegraphics[width=8.5cm]{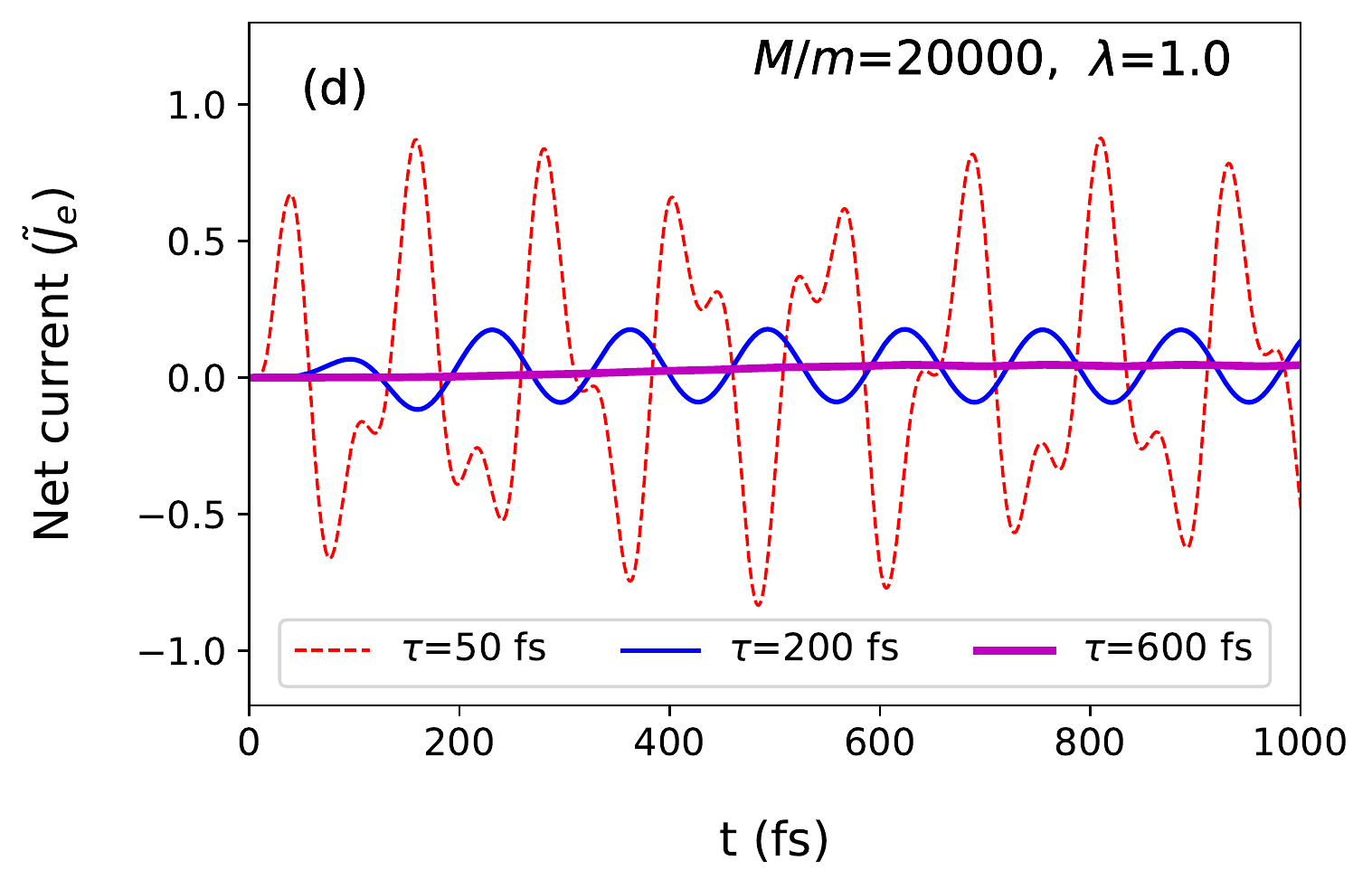}
   \caption{\label{current_M} The populations of $|k,l\rangle$ basis consisting
  of $|0,1\rangle$, $|1,0\rangle$ and $|2,-1\rangle$ states in the BH expansion
  of the three vibronic states  $j_1=u$ (a), $v$ (b) and $w$ (c) as function
  of nuclear electron mass ratio ($M/m$), at the full coupling limit(i.e. $\lambda=1.0$). In (d), the stationary and oscillatory currents generated by the action of RCP with different durations ($\tau$), as function of time, are shown for $M/m$=20000 at $\lambda=1.0$.}
  \end{figure*}
 The effect of the electronic-nuclear coupling is studied by multiplying the
  coupling constants in Eq.~(\ref{V2}) by a factor $\lambda$ in the $[0,1]$ range.
  A long RCP in the full coupling case, Fig.~\ref{current_l}(c), populates
  exclusively the $1_1$ state. It results in a stationary current
  $\tilde{J}_e(t)$ limited to about $0.4$ after the pulse, which exactly
  coincides with the $Y_{1_1}$ value.
  An admixture of the $1_{-1}$ state would reduce the amount of current with
  respect to this theoretical maximum.
  Thus, as expected, the RCP pulse couples exclusively to the $q=+1$ subspace and
  achieves the maximum ring current permitted for the $0_0\to 1_1$ transition.
  As the pulse becomes shorter, a superposition of the $0_1$, $1_1$ and $2_1$
  states is obtained instead, leading to an oscillatory current.
  The oscillatory current
  generated at the full coupling limit (cf. Fig. \ref{current_l}(c))
  reveals changes in the direction of the ring currents in time, which
  originate from the joint contribution of the basis states
  $|0,1\rangle$ and $|2,-1\rangle$ to the vibronic states $1_1$ and $2_1$.
  Going back to the equivalent picture of
  the $\chi_l(\alpha,t)$ nuclear wavepackets in the
  BH representation of the wave function, the coupling terms in the VCH
  Hamiltonian (\ref{HV}) couple the various $l$ contributions and result in the
  current oscillations, in which
  the nucleus and the electron exchange angular momentum.
  
  As the electronic-nuclear coupling diminishes for $\lambda<1$
  two effects are seen (cf. Fig. \ref{current_l}(a-b)).
  First, the normalized net current after the pulse increases and
  approaches a value of 1. With a reduced vibronic coupling, the right-rotating
  component in the $1_1$ vibronic state becomes more dominant, meaning that the molecular factor
  approaches 1 as well.
  Second, the current generated by the shortest pulse becomes less oscillatory
  and for $\lambda=0$ (not shown) it does not oscillate at all. Even if the
  pulse has enough bandwith to overlap with various states of the $q=1$ branch,
  only the one with the largest $|0,1\rangle$ contribution has a significant
  transition dipole. Other vibronic states in this branch, in contrast, have
  a negligible
  Franck-Condon factor with the ground state as the vibronic coupling is
  reduced.

 Finally, we approach the question of the nature of the generated currents from the point of view of the
  mass ratio $M/m$ between the nucleus and the electron, which in the previous
  calculations was always set to 2000.
  As the nuclear mass increases, the contribution to each electronic circulation
  direction in each vibronic eigenstate $(0_1,1_1,2_1)$ of the $q=1$ block,
  shown by red and blue curves in
  Fig.~\ref{current_M}, tends to become equal, which brings the molecular factor,
  shown in magenta,
  towards zero in the $M/m\to\infty$ limit. Essentially, the basis states
  $|2,-1\rangle$ and $|0,1\rangle$ become energetically degenerate and contribute
  equally to the vibronic eigenstates.
  As we discussed before, a RCP/LCP only permits transitions into either the
  $q=1$ or $q=-1$ branch, respectively, so a \emph{stationary} ring
  current is not possible in this limit (see magenta line in Fig. ~\ref{current_M}(d)).
  Extending the argument to the usual $E\times e$ molecular Jahn-Teller scenario,
  i.e. with direct coupling between electronic states of
  $E$ symmetry, this corresponds to the situation where the nuclear
  wave function localizes at the bottom of the Mexican hat potential
  (see e.g. Ref.~\citenum{domcke2004conical}, Ch. 10) and has a
  negligible amplitude at the point in configuration space where the diabatic
  electronic states are degenerate.
  In this BO limit, no stationary ring currents can be generated by a laser
  interacting with the ground state.
  The BO limit should not be confused with fixing the nuclei at
  a specific geometry. Note that in the RRM and in the limit $M/m\to\infty$ the
  nucleus is still completely delocalized along its coordinate $\alpha$.

  Even though stationary ring currents are not possible
  in the $M/m\to\infty$,
  the generation of an oscillatory ring current remains possible
  using a short RCP/LCP pulse of sufficient coherent bandwidth to excite
  simultaneously the various states in the corresponding $q=\pm 1$ subspace, as
  can be seen in Eq.~(\ref{Josc}). The currents are shown
  in Fig.~\ref{current_M}(d) in red and blue curves. The vibronic eigenstates do not support
  a ring current individually, but the off-diagonal molecular factors
  $Y_{j_1,j_1^\prime}$ in Eq.~(\ref{Josc}) are not zero, as can be seen by inspecting
  the different values of the coefficients for the same basis state in
  Fig.~\ref{current_M}(a-c). All vibronic states with $q=1$ have some non-zero
  contribution of the basis state $|0, 1\rangle$ and therefore all transitions
  $0_0\to j_1$ are allowed.
  %
  %
  However, the time-averaged ring current vanishes (see Fig.
  ~\ref{current_M}(d)) because each vibronic state in the linear superposition
  carries the same weight for both electronic circulation directions. Hence, in
  this BO limit, it is not possible to permanently favor one circulation
  direction over the other.
  This may not be as dramatic as it sounds, as the mass ratio between nuclei and
  electrons is not infinite, and because the vibronic coupling is not infinitely
  strong either, thus in general allowing for the existence of vibronic states
  in the $q=\pm 1$ subspaces ($E_\pm$ vibronic states in molecules) that favor
  to some extent a specific circulation direction of the electrons.

  \section{Conclusions}
  We have analyzed in detail the generation of stationary and oscillatory
  ring currents in a model system featuring the necessary symmetry
  conditions to support ring currents that are required in more complex systems
  as well as a transparent vibronic coupling mechanism.
  In the spirit of the
  Moshinsky-Kittel~\cite{moshinsky_1968}, the Shin-Metiu~\cite{shi95:9285} or
  Loguet-Higgins models~\cite{Longuet-Higgins},
  the RRM Hamiltonian allows for both
  analytical considerations and exact numerical solutions,
  both within and beyond the BO approximation.
  Electron-nuclear interactions hinder the maximum amount of stationary ring
  current that can be achieved by applying long, circularly polarized pulses.
  We characterize this intrinsic limitation to the maximum ring current
  through state-dependent molecular factors that correspond to
  the difference between right- and left-circulation contributions of the electrons in a specific
  vibronic state. These molecular factors depend on the strength of the
  nonadiabatic coupling between electrons and nuclei.
  Short, circularly polarized pulses with sufficient bandwidth lead to
  oscillatory currents created by the superposition of vibronic eigenstates
  within the same symmetry block of the complete Hamiltonian.
  In the limit of a large nuclear-electronic mass ratio, where the BO
  approximation is fulfilled, degenerate pairs of vibronic eigenstates cannot
  support stationary currents and the oscillatory currents generated by RCP/LCP
  broad-band pulses average out to zero over time.
  The observations made on the RRM system are general and have
  consequences for applications of laser-generated ring currents in molecules
  and materials.

\section{Supporting material}
\noindent
See the SI for the derivation of the general expression for the oscillatory currents.

\section{Data available in article or SI}
\noindent
The data that supports the findings of this study are available within the article [and its SI].

\section{acknowledgement}
\noindent
   The authors thank Prof. Wolfgang Domcke for insightful comments on an earlier version of the manuscript and Prof. Horst Köppel for helpful discussions.
    The authors declare no conflicts of interest.

%

\clearpage

\end{document}